\documentstyle[twoside,fleqn,espcrc2,amsfonts,epsfig,graphics]{article}

\def\COMENTARIO#1{}        

\def\bbox#1{{ \mbox{\boldmath $#1$}} }
\def\sbbox#1{{ \mbox{\scriptsize\boldmath $#1$}} }
\def\RP#1{\mathbf{R}\mathrm{P}^#1}

\def\num#1{\hbox{$^{\rm #1}$}}

\title{Antiferromagnetism in four dimensions: search
       for non-triviality\thanks{
        Presented by J.~M.~Carmona.
        Partially supported by CICyT (Spain) \hbox{AEN93-0604-C03}, 
        AEN94-0218, and \hbox{AEN95-1284E}.
}}

\author{J.~L.~Alonso\address{Departamento de F\'{\i}sica Te\'orica,
	Facultad de Ciencias, \\
	Universidad de Zaragoza, 50009 Zaragoza, Spain},
	H.~G.~Ballesteros\address{Departamento de F\'{\i}sica Te\'orica I,
	Facultad de CC. F\'{\i}sicas, \\
	Universidad Complutense de Madrid, 28040 Madrid, Spain},
	I.~Campos\num{a}, 
        J.~M.~Carmona\num{a},
        J.~Clemente~Gallardo\num{a},
	L.~A.~Fern\'andez\num{b}, D.~I\~niguez\num{a},
	V.~Mart\'{\i}n-Mayor\num{b}, A.~Mu\~noz Sudupe\num{b},
	A.~Taranc\'on\num{a} and  C.~L.~Ullod\num{a}}

\begin{document}
 
\begin{abstract}
We present antiferromagnetism as a mechanism capable of modifying
substantially the phase diagram and the critical behaviour of statistical
mechanical models. This is particularly relevant in four dimensions, due
to the connection between second order transition points and the continuum
limit as a quantum field theory.


We study three models with an antiferromagnetic interaction: the Ising
and the O(4) Models with a second neighbour negative coupling, and
the $\RP{2}$ Model. Different conclusions are obtained
depending on the model.
\end{abstract}

\maketitle

An antiferromagnetic (AFM) system shows in general different critical 
behaviour 
from that of its ferromagnetic counterpart, as will be seen below.
 This can be used
to study the triviality problem in Quantum Field Theory, by computing
the critical exponents or the renormalized coupling, which is
expressed in terms of the Binder cumulant $U_L$ of the model as
\begin{equation}
g_R=\lim_{L\to\infty}g_L^{(4)}=\lim_{L\to\infty}(L/\xi_L)^d U_L,
\label{gr}
\end{equation}
and should be different from zero in a non-trivial theory.

One way by which one could hope to see new critical behaviour,
i.e. new universality classes, is by considering AFM
models. Actually, it has been shown in low dimensions that
antiferromagnetism can produce new critical behaviour. For example, in
the Ising Model we can introduce it by means of a second-neighbour
(related with higher derivatives in the action)
\COMENTARIO{ Yo suprimiria ``(related with higher derivatives in the action)''
ya que si se habla de derivadas se esta pensando en campos suaves,
mientras que si hay frustracion o desorden del estado fundamental no
tiene sentido hablar de derivadas } negative coupling, which produces
frustration in the system. This makes a new ground state to appear,
which in $2d$ gives a new transition line with new critical 
behaviour~\cite{Mor94}.

The $3d$ $\RP{2}$ AFM Model has also recently been studied \cite{rp2}. The
Hamiltonian is given by 
\begin{equation}
{\cal H}=- \beta \sum_{<i j>} ({\mbox{\boldmath $v$}}_i\cdot \mbox{\boldmath
$v$}_j)^2 ,
\label{Hrp2}
\end{equation}
where $v_i$ is a three component normalized vector. 
Here a negative coupling produces a {\it mixed} ground state, formed by
one sub-lattice oriented in a fixed direction and the other lying on the
plane perpendicular to it, which brings about a new universality class.

We have considered the following models in $4d$.

\section{AFM Ising Model}

This is the simplest AFM model: the Ising Model in a hypercubic
lattice with first- and second-neighbour couplings
\begin{equation}
{\cal H}=-\beta_1\sum_{\sbbox{n},\mu}
  \sigma_{\sbbox{n}}\sigma_{\sbbox{n}+\hat{\sbbox{\mu}}}-
  \beta_2\sum_{\sbbox{n},\mu<\nu}
  \sigma_{\sbbox{n}}\sigma_{\sbbox{n}+\hat{\sbbox{\mu}}+\hat{\sbbox{\nu}}}\ .
\label{Hising}
\end{equation}
It was simulated on a $V=L^4$ lattice with periodic boundary
conditions, using a Heat Bath update (clusters algorithms are not
effective in most of AFM models). The Finite Size Scaling (FSS) was
studied up to $L=24$~\cite{Alonso}.

This model turns out to have a rich phase diagram,
with different vacuum structures. There is a region of ferromagnetic order
(FM): \hbox{$\sigma_{\sbbox{n}}=\sigma_{\bf{0}}$} 
($\sigma_{\bf{0}}$ stands for a fixed spin); there
is also a phase, HPAF, which consists of a FM configuration
on a three-dimensional cube and AFM on the other direction $\mu$:
$\sigma_{\sbbox{n}}=(-1)^{n_\mu}\sigma_{\bf{0}}$; finally, in the PAF phase, 
we have FM on a two-dimensional plane and
AFM on the other two directions: 
\hbox{$\sigma_{\sbbox{n}}=(-1)^{n_\mu+n_\nu}\sigma_{\bf{0}}$},
where $\mu,\nu\,(\mu<\nu)$ can be any directions.
We define order parameters according to these structures.

\begin{figure}[t]
\epsfig{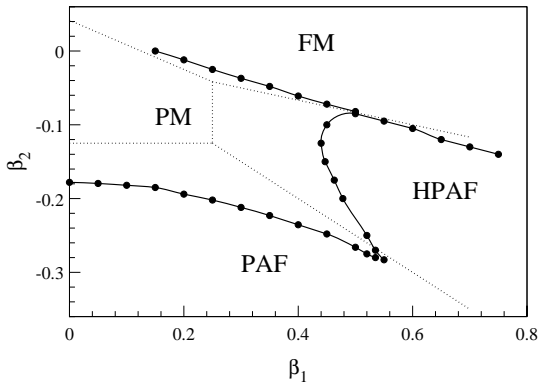}
\vglue -1cm
\caption{Phase Diagram of the $4d$ AFM Ising Model.}
\label{PhD}
\end{figure}          

The phase diagram is shown in Fig.~\ref{PhD}: PM means a disordered
phase and the dotted lines are the transition lines given by Mean
Field theory.  The transition line connected with the usual Ising
point ($\beta_2=0$) is second order with the same (classical)
exponents. The PM-HPAF line is first order.  And we find another line,
PM-PAF, which is disconnected from the Ising critical point. This
transition is clearly first order for values of $\beta_1$ greater than
$0.2$, but the behaviour is not so clear for $\beta_1<0.2$.

We produced a great amount of data at the points 
$\beta_1=0.1, 0.05, 0$. At $\beta_1=0$ the system decouples into two 
independent sublattices, known as $F_4$ lattices in the literature.
We found that the two peaks characteristic of
a first order transition are only distinguished at $\beta_1=0$, and with
the maximum size, $L=24$.

This AFM model teaches us that we have to be very careful, because first
order transitions with a very large correlation length can deceive you if
you do not go to sufficiently large lattice sizes to see the real
behaviour. With smaller sizes, apparent second order transitions arise,
and even false critical exponents can be measured. This system
behaves for small $L$ as a {\it weak} first order transition, and
presents the pseudo critical exponent $\alpha/\nu=1$, precursor of
a first order behaviour for large lattice sizes~\cite{weak}.

\section{AFM $\bbox{\RP{2}}$ Model}

This is the four-dimensional version of the model defined in (\ref{Hrp2}).
It was simulated by combining
Metropolis and Over-relaxed algorithms, up to $L=24$ lattice 
sizes~\cite{RP2D4}.

Two order parameters appear due to the special nature of the ground state,
one FM and the other AFM (or {\it staggered}). They are defined through 
tensors attached to every lattice site 
\begin{equation}
{\rm T}_i^{\alpha\beta}= v_i^\alpha v_i^\beta-
        \frac{1}{3}\delta^{\alpha\beta}\, ,
\end{equation}
as the normal or staggered sum of these tensors to the whole volume. 


In order to calculate critical exponents, we use
the FSS {\it ansatz} that allows to write
\begin{equation}
Q_O\equiv\frac{\left\langle O(2L,\beta)\right\rangle}
{\left\langle O(L,\beta)\right\rangle}=2^{x_O/\nu}
{F_O(\xi(2L,\beta)/2L)\over F_O(\xi(L,\beta)/L)}
\end{equation}
(except for scaling corrections), so that
\begin{equation}
\left.Q_O\right|_{Q_\xi=2}=2^{x_O/\nu}+\cdots,
\end{equation}
from where the critical exponent $x_O$ is readily extracted.

\begin{table}[b]
\caption{$g_R(L,\beta_c(\infty))$ for the $4d$ AFM $\RP{2}$ Model.}
\begin{tabular*}{\hsize}{@{\extracolsep{\fill}}rl}\hline
$\qquad\qquad L$&	$g_R$	                \\  \hline\hline
8	&	3.16 (3)	 	\\
10	&	2.84 (4)		\\
12	&	2.61 (5)		\\
16	&	2.34 (4)		\\
20	&	2.08 (9)		\\
24	&	1.95 (13)\qquad\qquad		\\  \hline
\end{tabular*}
\label{TABGR}
\end{table}

The model has a second order transition at \hbox{$\beta\approx-1.34$}, its
critical exponents being almost gaussian, the differences likely due to 
logarithmic corrections.
To know about the triviality of the theory we should 
therefore look at the renormalized coupling, defined by (\ref{gr}), at
the critical temperature. As can be seen in Table~\ref{TABGR}, for
the staggered sector the $g_R$
gets smaller values as the lattice size grows. We obtain a good fit to
$1/\log L$ without a constant term, so that the theory seems to be trivial.
However, in the
ferromagnetic sector, $g_R$ takes a negative, rather stable, value.
This might be an indication of a non-trivial limit for this theory.

\section{AFM O(4) Model}

We consider a system of spins 
\{$\bbox{\Phi}_{\sbbox{n}}$\} taking values in the hyper-sphere
$\mathrm{S}^3 \subset
\mathbf{R}^4$ and placed in the nodes of a hypercubic lattice. 
The interaction is defined by the Hamiltonian:
\begin{equation}
{\cal H} = - \beta_1 \sum_{\sbbox{n},\mu}
        {\bf \Phi}_{\sbbox{n}} {\bf \Phi}_{\sbbox{n}+\hat{\sbbox{\mu}}}
    - \beta_2 \sum_{\sbbox{n},\mu<\nu}
        {\bf \Phi}_{\sbbox{n}} {\bf \Phi}_{\sbbox{n}+\hat{\sbbox{\mu}} +
 \hat{\sbbox{\nu}}}\ .
\label{ACCION}
\end{equation}                   
\COMENTARIO{Quiza lo de que se trabaja en un reticulo hipercubico se
podria integrar y no decirlo 3 veces.}  The phase structure is found
to be qualitatively very similar to that of the Ising Model of
Fig.~\ref{PhD}, but now we have a continuum symmetry.  We did Monte Carlo
simulations in lattices ranging from $L=6$ to $L=24$ \cite{O4}. The
update method was a combination of Over-relaxed and Heat-Bath
algorithms, being $z\approx 1$.

In the O(4) AFM Model, the whole phase transition line
PM-PAF seems to be second order. 
We computed the critical
exponents on this line at $\beta_1 = 0$ coupling ($F_4$ lattice) 
with FSS techniques.
The results are quoted in Table~\ref{TABLE_EXPO}.
The correlation length exponent, $\nu$, is very close to the Mean 
Field prediction, namely $\nu =0.5$. 
Although we have not obtained a reliable measure of the
specific heat exponent our results point to $\alpha/\nu<1$
and disagree with the expected behaviour for a weak first order
transition.

\begin{table}[h]
\caption{Critical Exponents of the $4d$ AFM O(4) Model.}
\begin{tabular}{cccc}\hline
Lattice sizes  &$\gamma / \nu$  &$\beta / \nu$   &$\nu$  \\  \hline \hline
$6/12$                    &2.417(3)    &0.791(4)  &0.474(10) \\  
$8/16$                    &2.403(3)    &0.792(6)  &0.483(8)  \\  
$10/20$                   &2.410(2)    &0.790(4)  &0.471(6)  \\ 
$12/24$                   &2.403(5)    &0.797(5)  &0.483(7)  \\  \hline
\end{tabular}
\label{TABLE_EXPO}
\end{table}

\begin{figure}
\vglue 0.3cm
\epsfig{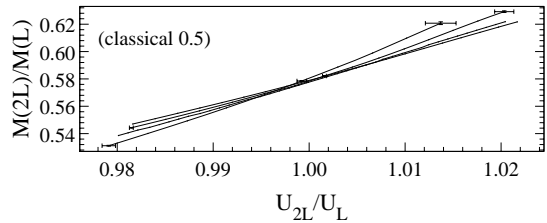}
\vglue -1cm
\caption{Quotients to obtain $\gamma/\nu$ (O(4)).}
\label{gamma}
\end{figure}

From our $\gamma/\nu$ estimation (see Fig.~\ref{gamma})
(or $\beta/\nu$ using the hyper-scaling relation) the exponent $\eta$
associated with the anomalous dimension of the field is $\eta\approx-0.4$.
This fact itself would imply the non-triviality of the theory, but
bearing in mind the results of section~1, transitory behaviour cannot 
be discarded. However, the stability of our
measures of $\gamma/\nu$ for the different lattice sizes,
(more than a hundred of standard deviations from
$\eta=0$), makes this rather unlikely.
We also measured the Binder cumulant at the critical point,
finding that it stays almost constant when increasing the lattice
size, which points to a
non-zero value of the renormalized constant in the thermodynamical
limit.                                                              

\end{document}